\documentclass[lettersize,journal]{IEEEtran}
\usepackage{amsmath,amsfonts}
\usepackage{cite}
\usepackage{algorithmic}
\usepackage{array}
\usepackage[caption=false,font=normalsize,labelfont=sf,textfont=sf]{subfig}
\usepackage{textcomp}
\usepackage{stfloats}
\usepackage{url}
\usepackage{verbatim}
\usepackage{graphicx}
\usepackage[numbers,sort&compress]{natbib}
\def\BibTeX{{\rm B\kern-.05em{\sc i\kern-.025em b}\kern-.08em
T\kern-.1667em\lower.7ex\hbox{E}\kern-.125emX}}
\usepackage{balance}
\begin{document}
\title{Influence of noise on Josephson junctions dynamics with a BCS theory-based model}
\author{L. Iwanikow, P. Febvre, \textit{Senior Member, IEEE}
\thanks{This work is supported by the Agence de l'Innovation de Défense (AID), the Centre National d'Etudes Spatiales (CNES) and the Office of the Director of National Intelligence (ODNI), the Intelligence Advanced Research Projects Activity (IARPA), via the U.S. Army Research Office Grant W911NF-17-1-0120.}}

\markboth{Submitted to ArXiv – 27 September 2023 - Revised on 10 November 2023}%
{Influence of noise on Josephson junctions dynamics with a BCS theory-based model}

\maketitle

\begin{abstract}
We developed a weak-linked Josephson junction time-domain simulation tool based on the Bardeen-Cooper-Schrieffer (BCS) theory to account for the electrodynamics of Cooper pairs and quasiparticles in the presence of thermal noise. The model, based on Werthamer and Harris formalisms, allows us to describe the behavior of Josephson junctions for various current and/or voltage input waveforms, analog or digital, and for any operating temperature below the critical temperature of the superconducting materials. We show a good agreement between experimental and simulated IV curves of Josephson junctions, as well as a relative increase of the grey zone width of an SFQ balanced comparator between 4 and 20$\%$ due to the presence of quasiparticles, for McCumber parameters comprised between 0.1 and 1, respectively.

\end{abstract}


\section{Introduction}

\IEEEPARstart{W}{ith} the rapid development of analog and digital superconducting electronics it is important to be able to determine accurately the expected performance of a circuit from the design stage. Indeed, one of the main challenges for the integration of complex superconductor-based circuits is to evaluate precisely their operation margins and yield through a good understanding of their behavior for any condition of operation. For instance the performance of balanced comparators subjected to microwave Single-Flux-Quantum (SFQ) clocks for digital circuits in presence of thermal noise is of high interest as it is a fundamental building block of superconducting digital electronics \cite{Filippov_1995}, analog-to-digital converters and sensitive detectors for single photon detection in quantum applications \cite{Steinhauer_2021}.

The Josephson junction \cite{Josephson_1962,Josephson_1974}, building block of superconducting electronics, is a highly non-linear device. The current flowing through it is based on the tunneling of two kinds of charge carriers through the insulated barrier between the two superconducting electrodes: Cooper pairs flowing without resistance, and quasiparticles. The latter, which result from the breaking of Cooper pairs at non-zero temperature, are subjected to the internal resistance of the barrier.

To this end, several time-domain simulators for Josephson junctions have been developed, such as JSIM \cite{Fang_1989}, PSCAN \cite{Polonsky_1991}, WRSpice \cite{WRSpice} or JoSIM \cite{Delport_2019}. With this in mind, we have developed a C++ time-domain simulator \cite{Iwanikow_2023} based on the results of BCS theory \cite{Bardeen_1957} to assess the impact of quasiparticles, whose effects are often neglected, and thermal noise on a circuit's dynamics and performance, especially, but not exclusively, for SFQ applications.

The model we implemented differs from the usual resistively and capacitively shunted junction (RCSJ) model as it takes into account the presence of quasiparticles on the behavior of the Josephson junction through the analysis of of the densities of states of Cooper pairs and quasiparticles, described by the BCS theory. Thermal effects influences the behavior of circuits through the proportion of quasiparticles created by Cooper pairs breaking, and through Johnson noise on the resistive parts of the circuit under study.

\section{Method}
The time-domain simulator of Josephson junctions is based on the description of the densities of states of Cooper pairs and quasiparticles described by BCS theory. The formalism developed by Werthamer \cite{Werthamer_1966}, then taken up and generalized by Harris \cite{Harris_1974,Harris_1975}, was used to deduce the dynamics of these charge carriers from their density of states. The time-domain response of a Josephson junction is then a function of the waveform applied to it:

\begin{equation}\label{eq:timeresponse}
\begin{aligned}
I(t) {} & = \Im \biggl\{ W(t) \int_{-\infty}^{+\infty} J_{p}(t-\tau)W(\tau) \,d\tau  \\
 & + W^{*}(t) \int_{-\infty}^{+\infty} J_{qp}(t-\tau)W(\tau) \,d\tau \biggr\} \\
 & = I_{p}(t) + I_{qp}(t)
\end{aligned}
\end{equation}
where the phase factor is $W(t) = \exp(j\varphi(t) /2)$, and $\varphi$ is the Josephson junction phase. $J_{p}$ and $J_{qp}$ describe the electrodynamics of the Cooper pairs and quasiparticles respectively, and are assimilated as sums of complex exponential functions.

\begin{figure}[!t]
\centering
\includegraphics[width=\linewidth]{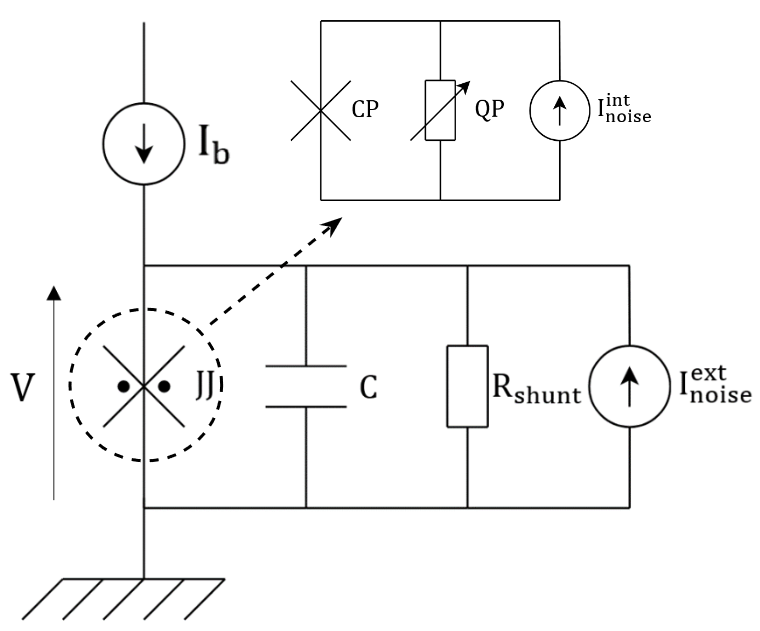}
\caption{Example of a Josephson junction equivalent circuit with a Norton noise current source. In this model, both Cooper pairs and quasiparticles are accounted in the JJ model, while the resistor \textit{R\textsubscript{shunt}} represents an external shunt. This shunt can be set to $\infty$ for non-shunted Josephson junctions. The barrier and shunt resistors each have their own noise current source.}
\label{fig:JJ_circuit}
\end{figure}

Moreover, the presence of thermal noise within a circuit, whether analogue or digital, can have a significant impact on its performance, such as a change of the apparent critical current of the Josephson junction, or the appearance of a stochastic behavior for junctions close to their switching point. It is therefore important to take these effects into account in order to predict the behavior of circuits under real conditions. An analytical work in presence of noise was performed by Ambegaokar \cite{Ambegaokar_1969} for some sets of parameters. In this work we performed a numerical analysis based on the above-mentioned formalism, taking the thermal noise into account. As such, we consider a thermal noise current generated by a Norton current source, described by a Gaussian distribution with a standard deviation per bandwidth unit $\sigma = \sqrt{4k_{B}T/R}$ for each resistance $R$ in the considered circuit, as shown in Fig. \ref{fig:JJ_circuit}.

Numerically, we take thermal noise into account as two separate contributions: we make the assumption that the current due to quasiparticles is, for each time $t$, equal to $I_{qp}(t) + I^{int}_{noise}(t)$, and that the shunt current can be described by $I_{shunt}(t) = V(t)/R_{shunt} + I^{ext}_{noise}(t)$, where $I^{int}_{noise}$ and $I^{ext}_{noise}$ are shown in Fig. \ref{fig:JJ_circuit}.

\newlength\imagewidth
\newlength\imagescale

For a current-biased junction, a Newton-Raphson (NR) algorithm has been used to determine the voltage evolution across the Josephson junction in the time domain. A multidimensional NR method must be set to describe circuits made of several junctions and components, such as inductors, resistors or capacitors.  

\section{Current-voltage characteristics}

Current-voltage \textit{IV} curves can be obtained by doing time-domain simulations with a varying control current. To do this, we perform a time simulation for a constant value of the bias current $I_{b}$, and determine the corresponding time evolution of the voltage across the simulated junction. We then average over time the value of this voltage, which gives one point on the \textit{IV} curve \cite{Razmkhah_2020}. We repeat this procedure for different values of the bias current by scanning current values in the increasing and then in the decreasing direction, in order to account for a potential hysteresis in the \textit{IV} curve, which is expected for junctions whose McCumber parameter $\beta_{C}=\frac{2\pi}{\Phi_0}R_{n}^{2}CI_{C}>1$, where $I_{C}$, $C$ and $R_{n}$ are respectively the junction critical current, capacitance and normal resistance, while $\Phi_0$ is the quantum of magnetic flux.

\begin{figure}[!t]
\centering
\includegraphics[width=\linewidth]{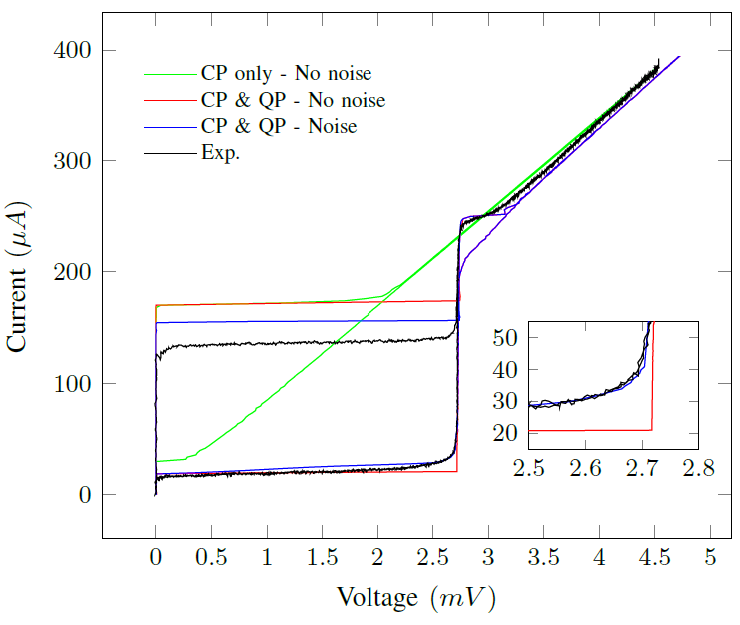}
\caption{Illustration of quasiparticles and thermal noise effects on niobium-based Josephson junctions at \textit{4.2 K} with $I_{C}$ \textit{= 172 $\mu$A}, $V_{g}$ \textit{= 2.74 mV}, $R_{n}$ \textit{= 13.5 $\Omega$}, $R_{subgap}$ \textit{= 95 $\Omega$} and \textit{C = 76.2 fF} compared to an experimental IV curve of a Josephson junction fabricated by the QUFAB Foundry in Japan (wafer HSTPA006 No.1 Chip E5), with a critical current density $j_{C}$ close to \textit{10 kA/cm²}.}
\label{fig:IV_analog}
\end{figure}

As a way to confirm or disprove the reliability of the simulator, we compared in Fig. \ref{fig:IV_analog} the results of experimental \textit{IV} curves to the ones resulting from our simulations.

We can see that the presence of thermal noise can cause the Josephson junction to switch at a bias current lower than the critical current. This results in a reduction in the apparent critical current of \textit{15 µA} (about $10\%$) with a standard deviation of \textit{6 µA}. There is also a smoothing of the return current below the gap, with a good agreement with the experimental curve, as shown in the inset of Fig. \ref{fig:IV_analog}.
Nevertheless we can still observe a difference of critical currents between simulated and experimental \textit{IV} curves. This may come from the presence of a magnetic field due to a nearby trapped flux during the measurement, which is not taken into account in our simulation tool, or from extra unfiltered noise brought by the measurement system which makes the junction switch earlier than expected. Moreover, the ratio $R_{n}I_{C}/V_{g} = (\pi/4) \, tanh(eV_{g}/4k_{B}T)$, where $V_{g}$ is the gap voltage, is only true for weak-coupling superconductors \cite{Golubov_1995}. We can also observe an hysteretic behaviour just over the gap on the simulated curves, which is not present in experimental measurements.

\section{SFQ balanced comparator}

\begin{figure}[!t]
\centering
\includegraphics[width=\linewidth]{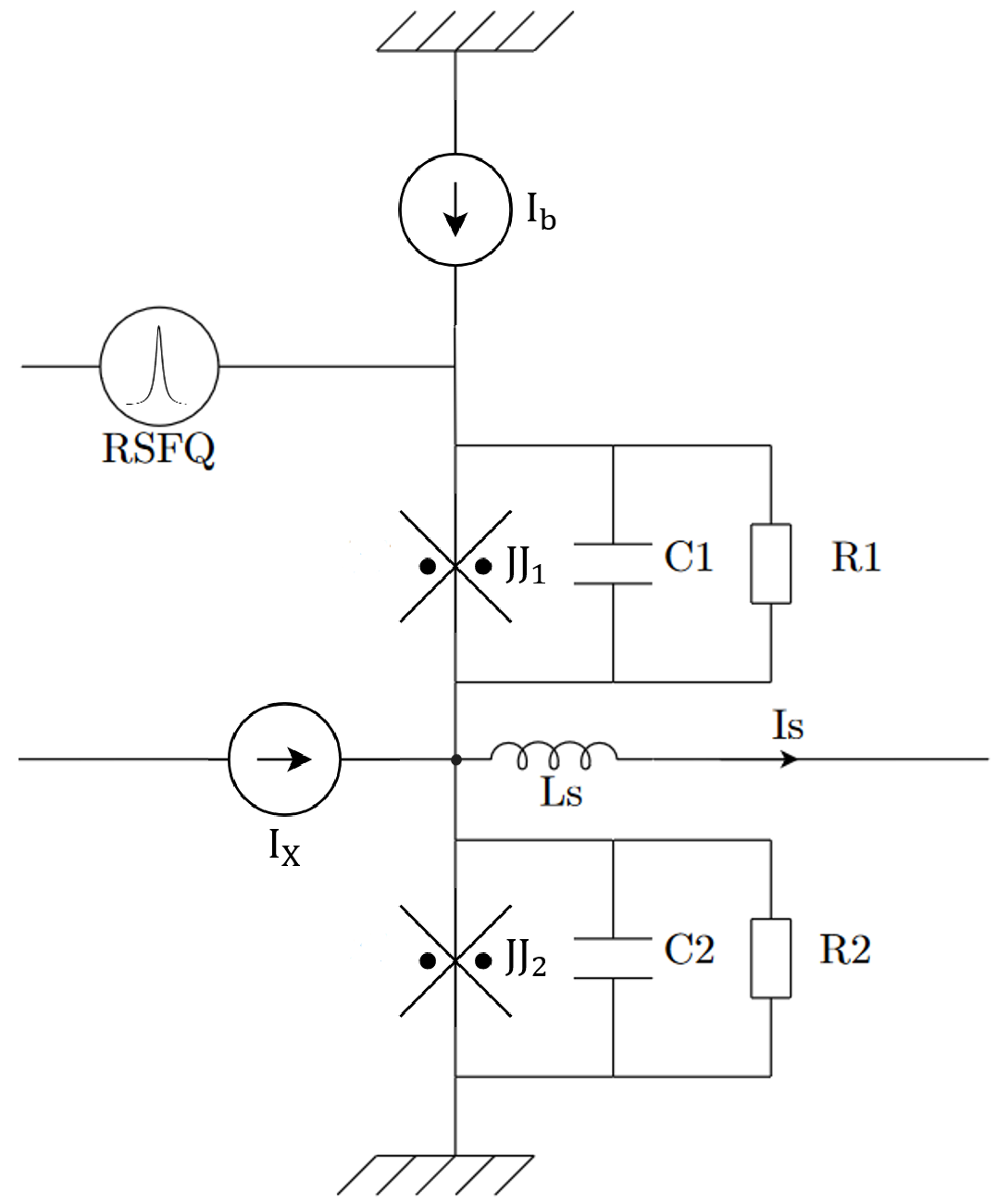}
\caption{Equivalent circuit of the simulated SFQ balanced comparator}
\label{fig:digital_comparator_circuit}
\end{figure}

The general time-domain framework shown before, set to simulate Josephson junction dynamics in the general case, can be used for instance to predict the IV curve of analogue Superconductor-Insulator-Superconductor (SIS) receivers in presence of microwave radiation \cite{Iwanikow_2023} or anticipate in a more general way spectral properties of Josephson devices through signatures on their \textit{IV} curves \cite{Divin_1999,Febvre_2000}. We point out here that the main advantage of time domain-based simulations is to compute the Josephson junctions dynamics in real time, which is especially useful for digital logic and any other effect based on dynamics taking place at short time scales, at the cost of more computing-intense simulations though.
To illustrate this point we took the example of a crucial brick of superconductor logic for analogue-to-digital conversion, SFQ and Adiabatic Quantum Flux Parametron (AQFP) logic cells: the Josephson balanced comparator \cite{Filippov_1995,Filippov_1999,Terai_2004,Yamanashi_2017}, as shown in Fig. \ref{fig:digital_comparator_circuit}. The values of electrical parameters of both JJ\textsubscript{1} and JJ\textsubscript{2} are given in Table \ref{tab:comparator_parameters}. In this study, we set $I_{b} = 125$ \textit{µA} and $L_{S} = 2.2$ \textit{pH}. Parameters where chosen such as $\beta_{C} = 1$ and $R_{n}I_{C} = 0.256$ \textit{mV}, with a critical current density $j_{C} = 10$ \textit{kA/cm²}.
The method described in this paper gives the possibility to assess the effects of Cooper pairs on the comparator properties, such as the commutation rate, in absence or in presence of quasiparticles, in order to evaluate the approximations usually done in the past, where the influence of quasiparticles was often neglected.

\begin{table}
\begin{center}
\caption{Electrical parameters of the simulated balanced comparator.}
\label{tab:comparator_parameters}
\begin{tabular}{| c | c | c | c |}
\hline
JJ\textsubscript{1} & Value & JJ\textsubscript{2} & Value\\
\hline
$I_{C_{1}}$ & \textit{164 µA} & $I_{C_{2}}$ & \textit{200 µA}\\
\hline
$R_{n_{1}}$ & \textit{13.0 $\Omega$} & $R_{n_{2}}$ & \textit{10.7 $\Omega$}\\
\hline
$C_{1}$ & \textit{0.82 pF} & $C_{2}$ & \textit{1.0 pF}\\
\hline
$R_{1}$ & \textit{1.56 $\Omega$} & $R_{2}$ & \textit{1.28 $\Omega$}\\
\hline 
\end{tabular}
\end{center}
\end{table}

\begin{figure}[!t]
\centering
\includegraphics[width=\linewidth]{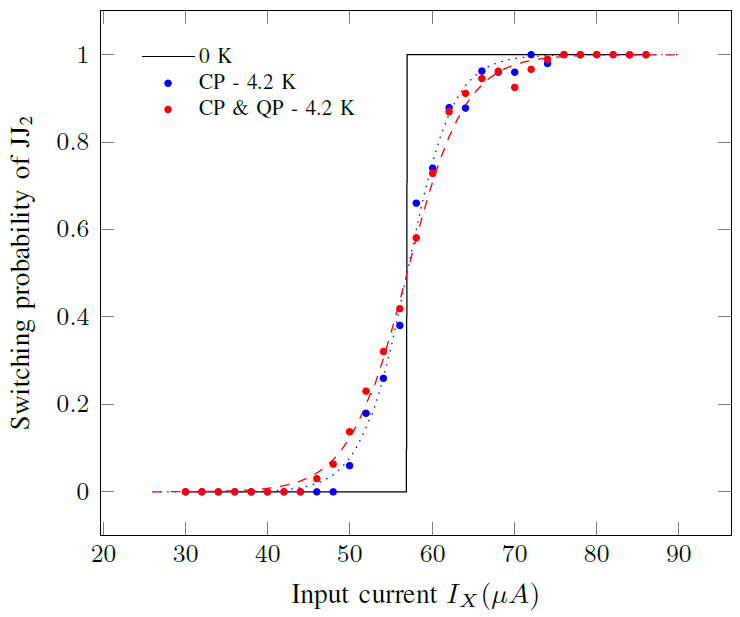}
\caption{Evolution of the probability of commutation of JJ\textsubscript{2} with the input current of a niobium-based Josephson junction at \textit{4.2 K} in the comparator circuit subjected to a \textit{50 GHz} SFQ voltage pulse train with and without quasiparticles effects. The black line represents the ideal case in the absence of thermal noise, while the blue and red markers show the simulation results while taking into account, respectively, the Cooper pairs only, then both Cooper pairs and quasiparticles. The dashed and dotted lines are the fitting results of the simulation with Eq. \ref{eq:probability_fit} for Cooper pairs with and without quasiparticles respectively.}
\label{fig:comparator_probability}
\end{figure}

The principle of such a comparator is to induce the commutation of the lower or upper Josephson junction, while the output signal is taken from only one of the two junctions, usually the lower one. The commutation of each of the two junctions depends on the value of the signal input $I_{X}$, which alters the current distribution in both JJ\textsubscript{1} and JJ\textsubscript{2}: for low values of $I_{X}$, the upper junction will be closer from its commutation state, which will make it switch instead of the lower junction, for each SFQ pulse delivered by the clock. However, when $I_{X}$ is high enough, the biasing current of the upper junction will be lower, which will favour the commutation of the lower junction.
Since the temperature is not at absolute zero, the probability of commutation of a Josephson junction biased with a current $I_{b}$ is not strictly 0 when $I_{b} < I_{C}$, or 1 when $I_{b} > I_{C}$. Instead, a grey zone exists around $I_{b} \approx I_{C}$ where the probability of commutation of the Josephson junction is between 0 and 1 when subjected to an SFQ pulse in this studied case of the SFQ comparator, as  shown in Fig. \ref{fig:comparator_probability}.

To study the range of use of the comparator, the probability of commutation of one of the junctions was calculated by counting its number of commutations between initial and final instants $t_0$ and $t_f$, for a range of values of $I_{X}$ at a set clock frequency $f_{clock}$. This number is then divided by the number of periods of the SFQ clock signal to obtain the probability. In practice and for physical reasons, it is easier to obtain this probability by working on the superconducting phase. Indeed, each SFQ pulse derives from a variation of phase $\Delta\varphi = 2\pi$ such that the probability of commutation is:
\begin{equation}\label{eq:probability_determination}
 P = \frac{\varphi(t_{f}) - \varphi(t_{0})}{2\pi  f_{clock} (t_{f} - t_{0})}
 \end{equation}
 
The probability of commutation of the other junction of the comparator is equal to $1 - P$.

The evolution of the probability as a function of the DC input current $I_{X}$ can be approximated by the equation:   

\begin{equation}\label{eq:probability_fit}
P_{2}(I_{X}) = 1 - P_{1}(I_{X}) = \frac{1}{2}[ 1 + erf(\sqrt{\pi} \frac{I_{X} - I_{thr}}{\Delta I})]
\end{equation}
where $erf$ is the error function.

To obtain a relevant comparison of the evolution of probability as a function of input current, we have kept the same value for the $RIc$ product for each $\beta_C$, with $R = (R_{n}R_{shunt})/(R_{n}+R_{shunt})$ and $R = R_{shunt}$ for the simulations with and without quasiparticles contribution, respectively.

The procedure used to obtain the data shown in Fig. \ref{fig:comparator_probability} is then applied for various clock frequencies between \textit{10 GHz} and \textit{100 GHz} to estimate the effect of the SFQ clock.

\begin{figure}[!t]
\centering
\includegraphics[width=\linewidth]{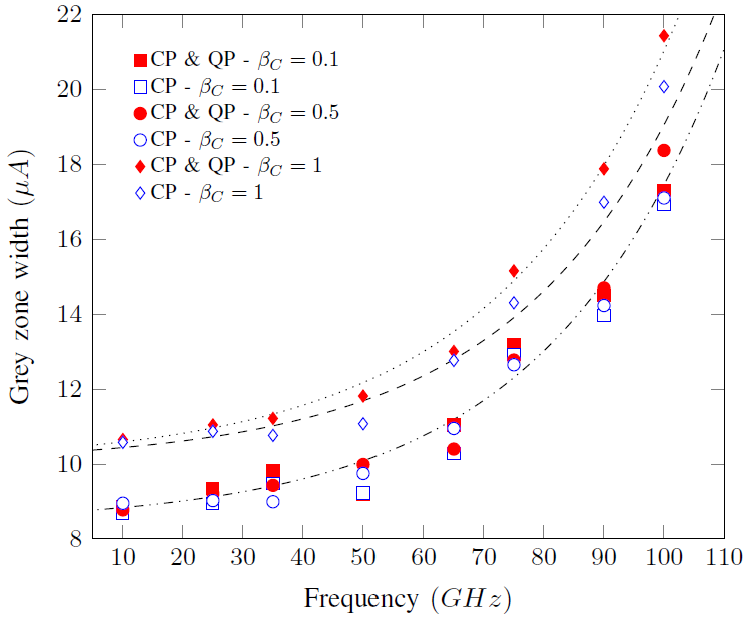}
\caption{Study of the effect of the SFQ clock frequency on a comparator circuit with or without taking the quasiparticles into account at 4.2 K, for different values of the Stewart-McCumber parameter. Markers are the results obtained through our simulation tool. Dotted and dashed lines are a guide for the eye for the evolution at $\beta_C=1$ of the grey zone with and without quasiparticles respectively, while the dot-dashed line is a guide for the eye for $\beta_C < 1$.}
\label{fig:comparator_widthgreyzone}
\end{figure}

As shown in Fig. \ref{fig:comparator_widthgreyzone}, the width of the grey zone increases with clock frequency, decreasing the sensitivity of the comparator, as also known from \cite{Filippov_1999}. We are also able to see the effect of quasiparticles on the dynamics of the circuit: their impact is minimal for low frequencies but increases with clock frequency. For example, taking quasiparticles into account increases by 18 \% the width of the grey zone for a  \textit{50 GHz} clock signal at \textit{4.2 K} for $\beta_{C} = 1$. As such, their impact may be critical to estimate operation margins, especially for high frequency applications.

In such circuits, Josephson junctions are usually moderately damped with $\beta_{C} = 1$ with a shunt resistor. A thorough analysis of the effect of Stewart-McCumber parameter $\beta_{C}$ on the performances of the comparator was done through the simulation of the circuit with altered values of the shunt resistors of JJ\textsubscript{1} and JJ\textsubscript{2} to decrease the value of $\beta_{C}$.

\begin{figure}[!t]
\centering
\includegraphics[width=\linewidth]{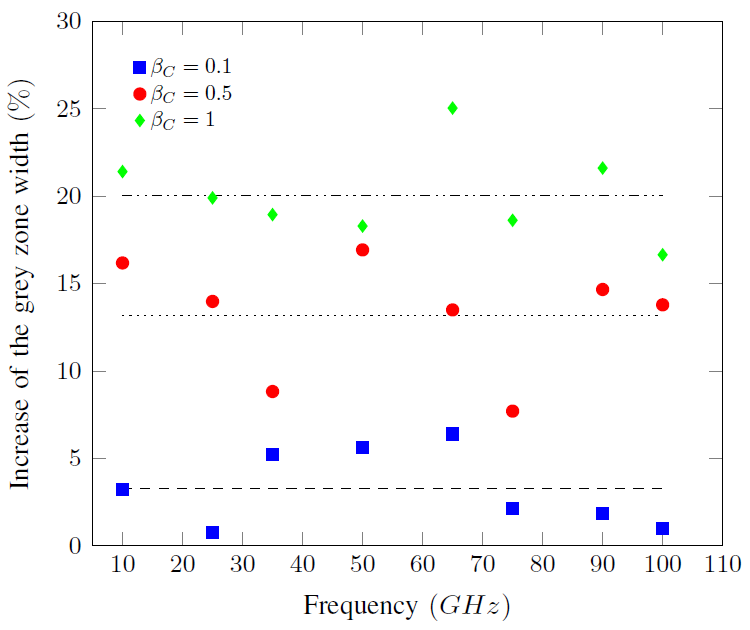}
\caption{Influence of the Stewart-McCumber parameter on the relative difference of the comparator grey zone induced by quasiparticles for clock frequencies ranging from 10 to 100 \textit{GHz}. Markers are the results obtained through our simulation tool, while dashed lines are a guide for the eye.}
\label{fig:mccumber_effect_QP}
\end{figure}

In Fig. \ref{fig:comparator_widthgreyzone}, we see that for overdamped Josephson junctions ($\beta_C < 1$), we obtain a better operation margin of roughly 15 \% on the whole clock frequency range. We may also note that the effect of quasiparticles on the operation margin seems less significant, which makes sense since the quasiparticles current path is shunted externally, leaving Cooper pairs to dominate the tunneling process. Fig. \ref{fig:mccumber_effect_QP} illustrates the impact of quasiparticles on operation margins for $\beta_C = 0.1, 0.5$ and $1$. If the impact seems to be steady on the studied range of clock frequencies for the simulated comparator, the effect of quasiparticles on the operation margins goes down from roughly 20 \% for $\beta_{C} = 1$ to 4 \% for $\beta_{C} = 0.1$.

A comparison with experimental measurements will be carried out in the near future to ensure the reliability of the simulations. However, the simulation model will have to be refined to take into account the contributions of the surrounding environment, such as external noise and parasitic elements.

\section{Conclusion}

The simulator allows us to qualitatively and quantitatively assess the effects of quasiparticles and thermal noise on the dynamics of Josephson junctions, for both analogue and digital applications. Comparisons between experimental measurements and simulations have been carried out on \textit{IV} curves for analogue applications and are in good agreement.
As such, this simulation tool can be of great help in the design of analog, digital or mixed-signal circuits. By describing more accurately the behavior of Josephson junctions by taking into account the role of quasiparticles, it can be used for example to determine more precisely the operation margins of logic functions in superconducting digital circuits, and improve the reliability of simulations during the design phase. Nevertheless further improvements are still needed, notably to account for effects of local magnetic fields.

\section*{Acknowledgments}

We want to thank Dr Mutsuo Hidaka who provided experimental data of IV curves associated with the fabrication of Josephson junctions made at the QUFAB Foundry in Japan with the HSTP technological process for prototyping quantum computer circuits. We are also very grateful to Thierry Robert, Jean-Luc Issler and Jean-François Penn for their continuous support.



\end{document}